\documentstyle[twocolumn,aps,psfig]{revtex}

\begin{document}
\draft

\twocolumn[\hsize\textwidth\columnwidth\hsize\csname @twocolumnfalse\endcsname

\title{Field induced transition of the $S=1$ 
antiferromagnetic chain with anisotropy
}

\author{T\^oru Sakai}
\address{
Faculty of Science, Himeji Institute of Technology, Kamigori,
Ako-gun, Hyogo 678-1297, Japan
}

\date{June 98}
\maketitle 

\begin{abstract}
The ground state magnetization process 
of the $S=1$ antiferromagnetic chain with 
the easy-axis single-ion anisotropy described by negative $D$ is
investigated. 
It is numerically found that 
a phase transition between two different gapless phases 
occurs at an intermediate magnetic field between the starting 
and saturation points of the magnetization for $-1.49<D<-0.35$. 
The transition is similar to the spin flopping, but 
it is second-order and not accompanied with any significant 
anomalous behaviors in the magnetization curve. 
We also present the phase diagrams in the $m$-$D$ and $H$-$D$ planes 
which reveal a possible re-entrant transition. 

\end{abstract}

\vskip2pc]
\narrowtext

\section{INTRODUCTION}
The one-dimensional (1D) antiferromagnets show various 
strange phenomena due to quantum effects. 
One of interesting examples appears 
in the low-temperature magnetization 
process of anisotropic antiferromagnets with an easy axis for the 
N\'eel order. 
In higher dimension including 2D 
the external magnetic field $H$ along the easy axis 
gives rise to a first-order phase transition
called spin flopping which results in a jump in the magnetization 
$m$ of the $S={1\over 2}$ Heisenberg antiferromagnet with the 
Ising-like anisotropic coupling.\cite{kohno} 
In 1D, however, 
the transition was shown\cite{yang} to be a second-order one 
described by the critical behavior $m\sim |H-H_c|^{1/2}$, 
where $H_c$ is the critical field. 
It implies that the discontinuity of the magnetization curve 
vanishes due to the strong quantum effect in 1D. 

The $S=1$ systems can have another-type easy-axis anisotropy 
described by $D\sum _j (S_j^z)^2 \quad (D<0)$ which would also 
yield the spin flopping in higher dimension.  
The 1D $S=1$ antiferromagnet,  
,even in the isotropic case, has 
a quite different feature from $S={1\over 2}$, 
which is characterized by a gap in the low-lying excitation 
spectrum, called Haldane gap (HG).\cite{haldane} 
Thus the magnetization process of the $S=1$ chain with the 
easy-axis anisotropy could include some interesting problems. 
In contrast to the easy-plane anisotropy ($D>0$),
the negative $D$ case has been investigated in few works,
partially because a noted quasi-1D $S=1$ antiferromagnet with negative
$D$ 
CsNiCl$_3$ was revealed to have the N\'eel order at
low temperatures,
which implies it doesn't have the Haldane gap.
However, it could be a good material to test the spin flopping
problem.

The ground state (GS) magnetization curve of the 1D $S=1$ isotropic 
antiferromagnet has already been given using the finite cluster 
calculation and size scaling
\cite{magnet}  
based on the conformal field theory(CFT).\cite{cft}
It has two critical fields $H_{c1}$ and $H_{c2}$ which denote  
the starting and saturation points the magnetization, respectively. 
$H_{c1}$ corresponds to the amplitude of HG, and 
GS has no magnetization and the exponentially decaying 
spin correlation for $H<H_{c1}$. 
The size scaling analysis also suggested that the transitions at 
$H_{c1}$ and $H_{c2}$ are second-order ones with the critical 
behaviors $m\sim (H-H_{c1})^{1/2}$ and 
$1-m\sim (H_{c2}-H)^{1/2}$, respectively. 
The further analysis\cite{size} on the excitation spectrum revealed that 
for $H_{c1}<H<H_{c2}$ the low-energy collective mode of the 
quasiparticles created by $S_j^+$'s is well described by the 
interacting fermion system called Luttinger liquid (LL)\cite{luttinger} 
which is characterized by the gapless charge and current 
excitations. 
In terms of the original spin system 
it implies that the excitation changing the uniform magnetization 
and the soft mode of the incommensurate spin density wave with 
the wave vector $2k_F=2\pi m$ are gapless for $0<m<1$.  

In the nonmagnetic GS of the 1D $S=1$ antiferromagnet 
the negative $D$ enhances the antiferromagnetic spin correlation 
along the easy axis and the system has the N\'eel order for 
sufficiently large $D$. 
The phenomenological renormalization group (PRG) 
analysis\cite{prg} up to the system 
size $L$=20 gave the Ising-like critical point $D_{c1}=-0.35 \pm 0.03$ 
which separates HG and N\'eel phases. 
For HG phase the magnetization process 
is expected to be qualitatively equivalent to the isotropic case.  
Note that even if HG vanishes 
due to the negative large $D$, 
$H_{c1}$ is still finite because the magnetic branches 
with $\sum S^z=\pm 1$ of the 
first excited triplet state go up in the energy spectrum 
due to the negative $D$, 
in contrast to the positive $D$.  
Since large negative $D$ excludes the state with $S_j^z=0$ at every site, 
each spin behaves like $S={1\over 2}$. 
In the limit $D\rightarrow -\infty$ 
the system is equivalent to the antiferromagnetic Ising model 
which shows only a trivial magnetization process described 
by only one step at $H_{c1}=H_{c2}$. 
The second-order perturbation of the exchange interaction, 
however, 
leads to the effective exchange process 
between $|-1,1\rangle$ and $|1,-1\rangle$
at adjacent sites. 
Thus the $S=1$ chain with large but finite negative $D$ 
is expected to be equivalent to the $S={1 \over 2}$ chain 
with Ising-like anisotropic couping. 
The equivalence seems to suggest that 
the $S=1$ chain with large negative $D$ shows quite similar 
magnetization process to the isotropic case; 
the two second-order phase transitions with the same critical 
behaviors as the isotropic case 
occur at $H_{c1}$, 
which corresponds to the Ising gap, and $H_{c2}$. 
The low-lying excitation of the system is also 
predicted to be described by  
LL for $H_{c1}<H<H_{c2}$.  
The characters of the gapless excitations, 
however, are different from the isotropic case. 
Since the state with $S_j^z=0$ leads to a large energy loss 
proportional to $D$, 
the quasiparticle of the gapless magnetic excitation should 
be described by $(S_j^+)^2$ instead of $S_j^+$. 
In addition the momentum of the soft mode is known\cite{luttinger} to 
be $2k_F=(1-m)\pi$, while it is $2k_F=2\pi m$ for the isotropic 
system. 
In terms of the interacting fermion systems, 
the difference in the soft mode is attributed to the change 
of the band picture which results from the pair creation 
due to the large on-site attraction made by $D$. 
Thus we distinguish 
LL phase that appears in the large negative $D$ 
region, denoted as LL2, 
from another LL phase including the isotropic case, 
denoted as LL1. 

In this paper 
we investigate the phase boundary between these two 
LL phases 
using the finite cluster calculation up to $L=20$ and 
some size scaling techniques. 
In addition we consider the possibility of the 
field induced phase transition, based on some phase diagrams, 
and the relation to the spin flopping. 

\section{MODEL AND NUMERICAL CALCULATIONS}
We start with the 1D $S=1$ antiferromagnetic Heisenberg Hamiltonian with
the single-ion anisotropy in a magnetic field 
\begin{eqnarray}
\label{ham}
&{\cal H}&={\cal H}_0+{\cal H}_Z, \nonumber \\
&{\cal H}_0& = \sum _j {\bf S}_j \cdot {\bf S}_{j+1}
+D\sum _j(S_j^z)^2, \\
&{\cal H}_Z& =-H\sum _j S_j^z, \nonumber
\end{eqnarray}
under the periodic boundary condition. 
We consider only the easy-axis anisotropy ($D\leq 0$) and 
the Zeeman term $\cal{H}_Z$ implies the applied magnetic field 
along the axis. 
For $L$-site systems,
the lowest energy of ${\cal H}_0$ in the subspace where
the eigenvalue of $\sum _j S_j^z$ is $M$
(the macroscopic magnetization is $m=M/L$) and 
the momentum is $k$, is denoted as $E_k(L,M)$.
In addition we define $E(L,M)$ as the lowest one among $E_k(L,M)$'s. 
Using Lanczos algorithm, we calculated $E_k(L,M)$
($M=0,1,2,\cdots,L$) for even-site systems up to $L=20$.

\section{SPIN CORRELATION FUNCTIONS AND SOFT MODES}
The two phases LL1 and LL2 are similar to the two planar phases 
in the nonmagnetic GS phase diagram in the $D$-$J_z$ plane, 
where $J_z$ is the $z$-component of the exchange coupling, 
obtained by a bosonization technique. 
\cite{schulz} 
It suggested that the phase boundary 
was Ising-like. 
Consulting the result from the bosonization, 
to clarify the features of LL1 and LL2 in GS, 
we consider the following three spin correlation functions: 
\begin{eqnarray}
\label{cor}
&C_1(r)&=\langle S_0^+S_r^- \rangle \sim (-1)^r r^{-\eta _1},\nonumber \\ 
&C_2(r)&=\langle (S_0^+)^2(S_r^-)^2 \rangle \sim r^{-\eta _2},\\ 
&C_z(r)&=\langle S_0^zS_r^z \rangle \sim \cos(2k_F) r^{-\eta _z},
\nonumber 
\end{eqnarray}
where we take only the most dominant term for each. 
$C_1(r)$ and $C_2(r)$ are associated with 
the gapless single-particle ($\delta M=\pm 1$) and   
two-particle ($\delta M=\pm 2$) excitations. 
$C_z(r)$ is related to the gapless soft mode with $2k_F$. 
It is expected that 
both $C_1(r)$ and $C_2(r)$ obey the power law shown in (\ref{cor}) in LL1, 
while only $C_1(r)$ decays exponentially in LL2  
because the pair excitation is gapless, 
but the excitation with $\delta M=\pm 1$ is massive in LL2. 
It is also predicted that 
the momentum of the soft mode corresponds to $2k_{F1}=2\pi m$ for LL1, 
$2k_{F2}=(1-m)\pi$ for LL2. 
Therefore, if we consider the single-particle excitation gap $\Delta _1$
and the soft mode gap with the momentum $2k_{F2}$ $\Delta _{2k_{F2}}$, 
the two gaps should cross at the phase boundary between LL1 and LL2. 
Because $\Delta _1=0$ and $\Delta _{2k_{F2}}\not=0$ in LL1, while 
$\Delta _1\not=0$ and $\Delta _{2k_{F2}}=0$ in LL2 in the thermodynamic
limit. 
%
%
\begin{figure}[htb]
\begin{center}
\mbox{\psfig{figure=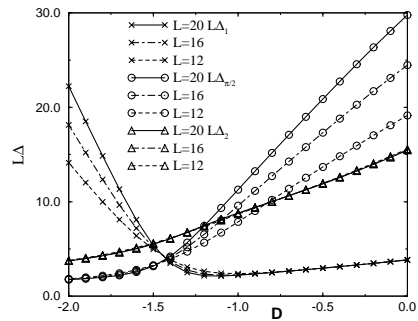,width=6cm,height=5cm,angle=-90}}
\end{center}
\caption{
Scaled gaps $L\Delta _1$ $(\times)$, $L\Delta _{\pi /2}$ $(\circ)$
and $L\Delta _2$ $(\triangle)$ are plotted vs $D$ for $L=20$, 16 and 12
with fixed $m=1/2$.
$L\Delta _2$ does not depend on $L$, which means it is always gapless
in the thermodynamic limit.
The crossing point of  $L\Delta _1$ and $L\Delta _{\pi /2}$ is
almost independent of $L$ and gives a good estimation of the
critical point as $D_c=-1.44 \pm 0.01 $.
\label{fig1}
}
\end{figure}

\section{PHASE BOUNDARY AND CRITICAL INDICES}
In order to investigate the phase boundary between LL1 and LL2 in GS  
with $m\not= 0$, 
we vary the parameter $D(<0)$ with fixed $m={M \over L}$. 
PRG is a good approach to study
on phase transitions, particularly with the Ising-like universality.
If we assume the critical behavior of the gap
$\Delta \sim |D-D_c|^{\nu}$,
the $L$-dependent fixed point $D_c(L)$ can be determined by
the PRG equation
\begin{eqnarray}
\label{prg}
L_{L+\delta L}\Delta (L+\delta L,D')=L\Delta (L,D),
\end{eqnarray}
and the $L$-dependent exponent $\nu(L)$ is also estimated by
\begin{eqnarray}
\label{nu}
\nu(L)=\ln \big[ {{L+\delta L}\over L} \big] \big/
\ln \big[ {{(L+\delta L)\Delta'(D_c(L),L+\delta L)}
\over {L\Delta'(D_c(L),L)}} \big],
\end{eqnarray}
where $\Delta '(D,L)$ is the derivative of $\Delta (D,L)$ with respect
to $D$.
The extrapolation of $D_c(L)$ and $\nu (L)$ to $\L\rightarrow \infty$
give $D_c$ and $\nu$ in the thermodynamic limit.

Based on the above argument,
the two gaps $\Delta _1$ and $\Delta _{2k_{F2}}$ are
taken for the order parameters.
The two gaps of the finite size systems are calculated by the forms
\begin{eqnarray}
\label{gap1}
&\Delta _1&=E_{\pi}(L,M+1)+E_{\pi}(L,M-1)-2E_0(L,M),  \nonumber \\
&\Delta_{2k_{F2}}&=E_{2k_{F2}}(L,M)-E_0(L,M),
\end{eqnarray}
where $M=mL$, and we restrict $M$ to an even number
because odd $M$ states cannot be the ground state in LL2.
The form of $\Delta _1$ is obtained by the sum of the two gaps
with $\delta M=+1$ and $-1$ of the total Hamiltonian ${\cal H}$,
to cancel out the Zeeman energies in both gaps.

\begin{figure}[htb]
\begin{center}
\mbox{\psfig{figure=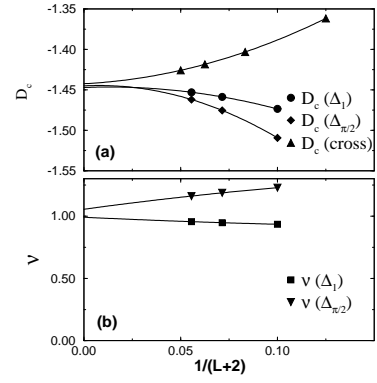,width=6cm,height=5cm,angle=-90}}
\end{center}
\caption{
(a)Some estimations of the critical point $D_c$ for $m=1/2$:
the $L$-dependent fixed points $D_c(L)$ for $\Delta _1$ (solid circle)
and $\Delta _{\pi /2}$ (solid diamond), and the crossing point of
the two gaps (solid triangle) plotted versus $1/(L+2)$.
The extrapolation by fitting of the quadratic function of $1/(L+2)$
gives $D_c=-1.45$, $-1.45$ and $-1.44\pm 0.01$, respectively.
(b)$L$-dependent exponents $\nu (L)$ for $\Delta _1$ (solid square)
and $\Delta _{\pi /2}$ (solid triangle down) plotted versus $1/(L+2)$.
The same extrapolation gives $\nu =1$ for both gaps with the error
less than a few percent.
They are consistent with the Ising like universality.
\label{fig2}
}
\end{figure}

For $m={1\over 2}$ where $2k_F={{\pi}\over 2}$ for LL2,
the scaled gaps $L\Delta _1$ and $L\Delta _{\pi /2}$ are plotted
versus $D$ for $L=$20, 16 and 12 in Fig. \ref{fig1}.
The two-particle excitation gap
\begin{eqnarray}
\label{gap2}
\Delta _2=E_0(L,M+2)+E_0(L,M-2)-2E_0(L,M),
\end{eqnarray}
is also plotted as $L\Delta _2$ in Fig. \ref{fig1}.
$L\Delta _2$ is almost independent of $L$ in the whole region,
which means the two-particle excitation is always gapless
because the size dependence of the gap is $\Delta \sim {1\over L}$
at a gapless point, as CFT predicted.
On the other hand,
the behaviors of $\Delta _1$ and $\Delta _{\pi /2}$ obviously
suggest the existence of a critical point $D_c$ between LL1 and LL2.
They imply that $\Delta _1$ ($\Delta _{\pi /2}$) is gapless (gapped)
in LL1, while gapped (gapless) in LL2.
In Fig. \ref{fig1} the $L$-dependent fixed points of $\Delta _1$ and
$\Delta _{\pi /2}$, in addition the crossing point of the two gaps
with fixed $L$, look almost independent of $L$.
The extrapolation of
the both $L$-dependent fixed points for $\delta L=4$
using the standard least-aquare fitting of the quadratic function
of ${1\over {L+2}}$
gives a common value $D_c=-1.45$,
which shows a good agreement with the extrapolated result of
the crossing point $D_c=-1.44$, as shown in Fig. \ref{fig2}(a).
Thus we take the result of the crossing point of $\Delta _1$
and $\Delta _{\pi /2}$ for the best estimation as $D_c=-1.44 \pm 0.01$.
The same extrapolation of the exponent $\nu$ gives the conclusion
$\nu =1$ for both gaps with the error less than a few percent,
as shown in Fig. \ref{fig2}(b).
It is consistent with the Ising universality.
%
%

\begin{figure}[htb]
\begin{center}
\mbox{\psfig{figure=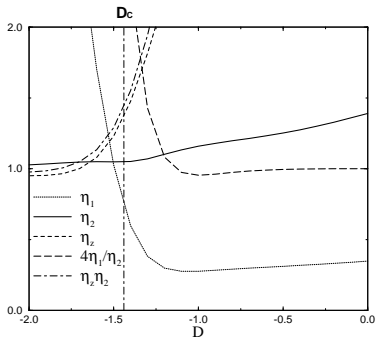,width=6cm,height=5cm,angle=-90}}
\end{center}
\caption{
Estimated exponents $\eta _1$ (dotted line),
$\eta _2$ (solid line) and $\eta _z$
(dashed line) are plotted versus $D$ for $m=1/2$.
The values of $4\eta _1 /\eta _2$ (long dashed line)
and $\eta _z \eta _2$ (dot-dashed line)
are also plotted to test the universal relations
$\eta_2 =4\eta$ for LL1 and $\eta_z \eta_2=1$ for LL2.
\label{fig3}
}
\end{figure}

For further test of the equivalency of the present phase boundary
between LL1 and
LL2 to the one between the two planar phases in the nonmagnetic
ground state phase diagram discussed above,
we estimate the exponents
$\eta _1$, $\eta _2$ and $\eta_z$ for $2k_{F2}$
in the form (\ref{cor}),
using the $L$ dependence of the gap associated
with each spin correlation function like
\begin{eqnarray}
\label{eta1}
\Delta \sim \pi v_s \eta /L,
\end{eqnarray}
where $v_s$ is the sound velocity which is 
obtained from the form 
\begin{eqnarray}
\label{vs}
v_s = {L\over {2\pi}}(E_{{2\pi}/L}(L,M)-E_0(L,M)) +O({1\over {L^2}}),
\end{eqnarray}
as CFT predicted. 
For $m={1\over 2}$ 
the estimated exponents $\eta _1$, $\eta _2$ and $\eta_z$, 
using the extrapolation of the size-dependent results from the form 
(\ref{eta1}) by fitting the quadratic function
of ${1\over L}$, 
are shown in Fig. \ref{fig3}. 
Except for the rapidly diverging behaviors of $\eta _1$ and $\eta _z$ 
near the critical point $D_c=-1.44$ 
where the size correction is too large to estimate them, 
all the exponents seem to approach the expected values 
$\eta _1={1\over 4}$, $\eta _2=1$ and
$\eta _z=1$, with $D$ approaching $D_c$. 
In addition the values of $4\eta _1 /\eta _2$ and $\eta _z \eta _2$ 
are plotted in Fig. \ref{fig3}, 
to check the universal relations 
$\eta_2 =4\eta _1$ for LL1 and $\eta_z \eta_2=1$ for LL2. 
The relations are revealed to be satisfied at least 
as far from the critical point as $|D-D_c|\gtrsim 0.3$. 
They are all consistent with the results of the bosonization for
the two phases in the nonmagnetic ground state.
Our estimation of
the central charge $c$ of CFT 
using the form of the size dependence of the
ground state energy
\begin{eqnarray}
\label{central}
{1\over L}E_0(L,M) \sim \epsilon (m) -{{\pi}\over 6}cv_s {1\over {L^2}},
\end{eqnarray}
results in $c=1$ at least sufficiently far from $D_c$, as expected 
for both LL1 and LL2, 
while $c=2$ just at $D_c$. 
It implies that
two Ising-like criticalities with $c={1\over 2}$ for each
described by $\Delta _1$
and $\Delta _{2k_{F2}}$ appear at $D_c$ on the background with $c=1$.
We also got the same result of every critical exponent
for $m={1\over 4}$ and ${3\over 4}$.
%
%

\section{PHASE DIAGRAMS}
\subsection{$m$-$D$ phase diagram}

\begin{figure}[htb]
\begin{center}
\mbox{\psfig{figure=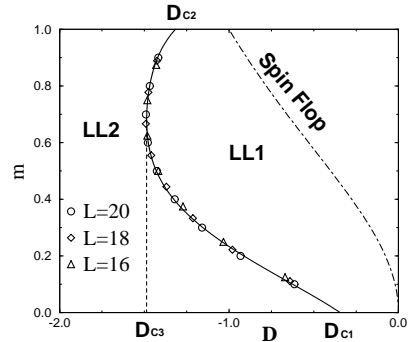,width=6cm,height=5cm,angle=-90}}
\end{center}
\caption{
$m$-$D$ phase diagram.
The boundary between LL1 and LL2 is determined by the polynomial curve
fitting (a solid line)
to the crossing points of $L\Delta _1$ and $L\Delta _{2k_{F2}}$
for various $m$'s with $L=20$ and 18,
which are almost independent of $L$.
The extrapolations of $D_c$ for $M=2 and L-2$
gave $D_{c1}=-0.35\pm 0.01$ and
$D_{c2}=-1.32 \pm 0.01$, respectively.
The above curve fitting yielded
$D_{c3}=-1.49$.
In addition the classical spin flopping line is shown as
a dot-dashed line.
\label{fig4}
}
\end{figure}

The crossing point  
of the two gaps $\Delta _1$ and $\Delta _{2k_{F2}}$, 
for various magnetizations that can be obtained by $m={M\over L}$ 
using even $M$ for $L$=16, 18 and 20, 
are plotted in the $m$-$D$ plane in Fig. \ref{fig4}. 
Since they give a common curve almost independent of $L$, 
we don't consider any size corrections except for the limits 
$m\rightarrow 0$ and 1. 
The extrapolation of the crossing point for various $L$ with 
fixed $M=2$ leads to the phase boundary in the limit $m=0+$ 
as $D_{c1}=-0.35 \pm 0.01$, 
which gives a good coincidence with 
the boundary between HG and N\'eel phases in the nonmagnetic GS 
derived from PRG.
A similar estimation yields $D_{c2}=-1.32 \pm 0.01$ in the limit $m=1-$.
Fitting a suitable polynomial to $D_c$ for $L=20$ and 18, 
and the extrapolated $D_{c1}$ and $D_{c2}$, 
the conclusive phase boundary is obtained as a solid line in Fig.
\ref{fig4}. 
It suggests that the phase transition from LL2 to LL1 
occurs in the magnetization process for $D_{c3}<D<D_{c1}$, 
where $D_{c3}$ is the minimum of $D_c$, as shown in Fig. \ref{fig4}, 
and $D_{c3}=-1.49$ is obtained by the fitting curve. 
In addition 
the phase diagram shows the existence of 
a re-entrant transition returning to LL2  
for $D_{c3}<D<D_{c2}$. 
%
%

\subsection{$H$-$D$ phase diagram}

\begin{figure}[htb]
\begin{center}
\mbox{\psfig{figure=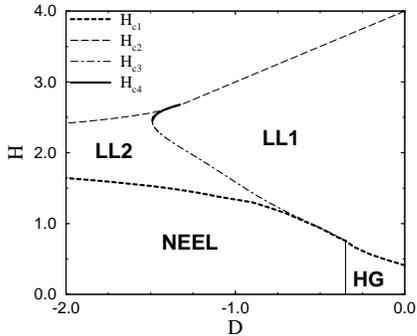,width=6cm,height=5cm,angle=-90}}
\end{center}
\caption{
$H$-$D$ phase diagram.
$H_{c3}$ (a dot-dashed line) and $H_{c4}$ (a thick solid line) were
obtained by plots of ($D_c$, $H_M^{(1)}$) for various $m$
with $L=20$ and 18, and
a curve fitting.
$H_{c1}$ (a thick dashed line) and $H_{c2}$ (a long dashed line)
were obtained by the Shanks transformation applied to $\Delta _1$ (LL1)
and $\Delta _2$ (LL2) corresponding to $m=0+$ and $m=1-$,
respectively.
The field induced transition from LL2 to LL1
occurs at $H_{c3}$ fon $-1.49<D<-0.35$,
and the re-entrant transition also exists for $-1.49<D<-1.32$.
\label{fig5}
}
\end{figure}

Since 
the excitation with $\delta M=\pm 1$ is gapless 
for $0<m<1$ in LL1, 
CFT predicts the asymptotic form 
\begin{eqnarray}
\label{field1}
\pm E(L,M\pm 1)\mp E(L,M) \sim H \pm A{1\over L}+o({1\over L}) 
\end{eqnarray}
where $A=\pi v_s \eta _1$. 
Thus the average of the two forms (\ref{field1}), 
canceling the dominant size corrections, gives  
a good estimation of $H$ for $m={M\over L}$ as 
\begin{eqnarray}
\label{h1}
H_M^{(1)}\equiv [E(L,M+1)-E(L,M-1)]/2 = H +o({1\over L}).
\end{eqnarray}
To get the phase boundary between LL1 and LL2 in $H$-$D$ plane, 
we calculate  
$H_M^{(1)}$ at $D_c$ for given $m$ with $L=20$ and 18, 
and plot ($D_c, H_M^{(1)}$) for various $m$'s. 
The curve obtained by 
the polynomial fitting to the plot,  is shown in Fig. \ref{fig5}, 
where $H_{c3}$ is the lower boundary and $H_{c4}$ is the upper. 
$H_{c1}$ is estimated by the Shanks transformation\cite{shanks} 
applied to $E(L,1)-E(L,0)$ for $L=12$ to 20, 
because the nonmagnetic GS is massive.  
In LL2, however, 
the correct magnetic field $H$ for $m$ cannot be obtained 
by $H_M^{(1)}$, 
because the excitation with $\delta M\pm 1$ has a finite 
gap proportional $D$. 
Thus we have to use another form 
\begin{eqnarray}
\label{h2}
H_M^{(2)}\equiv [E(L,M+2)-E(L,M-2)]/4 = H +o({1\over L}),
\end{eqnarray}
where only even $M$ is available. 
$H_{c1}$ and $H_{c2}$ at the boundary of LL2 are 
estimated by the Shanks transformation applied to 
$(E(L,2)-E(L,0))/2$ and $(E(L,L)-E(L-2))/2$, respectively. 
Fig. \ref{fig5} shows the completed $H$-$D$ phase diagram.
%
%

\section{MAGNETIZATION CURVE}

\begin{figure}[htb]
\begin{center}
\mbox{\psfig{figure=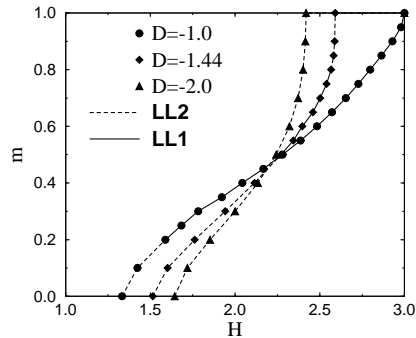,width=6cm,height=5cm,angle=-90}}
\end{center}
\caption{
Magnetization curve in the ground state for $D=-1.0$, $-1.44$ and
$-2.0$.
Solid and dashed lines correspond to LL1 and LL2, respectively.
The transition from LL2 to LL1 appears at $H_{c3}=-1.59$ for $D=-1.0$,
and at $H_{c3}=-2.27$ for $D=-1.44$.
The re-entrant transition occurs at $H_{c4}=-2.58$ for $D=-1.44$.
No significant anomalous behavior appears at any critical fields.
\label{fig6}
}
\end{figure}

Finally we investigate the behavior of the magnetization curve, 
in particular around the critical fields. 
If we assume the form around the critical field $H_c$ 
corresponding to the magnetization $m_c$ as 
\begin{eqnarray}
\label{delta}
m-m_c \sim (H-H_c)^{1/\delta}, 
\end{eqnarray}
the exponent $\delta$ can be estimated by the form\cite{exponent}
\begin{eqnarray}
\label{estimate}
\ln \big( {{f(L)}\over{f(L+2)}} \big) /\ln \big({{L+2}\over L})
\sim \delta \qquad (L\rightarrow \infty),
\end{eqnarray}
where $f(L)$ is given by
\begin{eqnarray}
\label{fm}
f(L)\equiv E(L,M_c+2)+E(L,M_c)-2E(L,M_c+1),
\end{eqnarray}
where $M_c=m_cL$.
For LL2 the forms need some modifications like $L+2 \rightarrow L+4$,
$M_c+2 \rightarrow M_c+4$ and $M_c+1 \rightarrow M_c+2$.
The method applied to $H_{c1}$ at $D=0$ (isotropic case) gave the
result $\delta =1.9 \pm 0.1$ consistent with
$\delta =2$ predicted by some effective Hamiltonian theories.
\cite{exponent}
Our present application of the method to $H_{c1}$ at $D=-2.0$ in LL2
yields $\delta =2.03 \pm 0.04$ which supports the equivalence
between the present model (\ref{ham}) and the anisotropic $S={1\over 2}$
chain, discussed above.
To estimate $\delta $ around $H_{c3}$,
we applied it to the phase boundary $D_c=-1.44$ for $m={1\over 2}$,
and got $\delta =0.9 \pm 0.3$ which suggests $\delta =1$.
It implies that the magnetization curve is linear without any
discontinuity around $H_{c3}$,
because any jump at $H_{c3}$ whoud make the estimated $\delta $
very small.
However,
the possibility of
a change in the derivative of the magnetization curve at $H_{c3}$ still
exists.
To test it,
we give the ground state magnetization curve for $D=-1.0$, $-1.44$ and
$-2.0$ in Fig. \ref{fig6},
using the form (\ref{h1}) in LL1 and (\ref{h2}) in LL2
for $L=20$.
The curve for $D=-1.0$ has $H_{c3}=-1.59$,
and the one for $D=-1.44$ has
$H_{c3}=-2.27$ and $H_{c4}=-2.58$.
In Fig. \ref{fig5} solid and dashed lines stand for LL1 and LL2,
respectively.
At least within this analysis,
no significant anomalous behavior can be detected around $H_{c3}$
or $H_{c4}$.
%
%

\section{DISCUSSIONS}

To consider the relation of the transition at $H_{c3}$ 
to the spin flopping,  
the line of the spin flopping in the classical limit with 
$({\bf S}_j)^2=1$, which is  $D=-2m^2/(1+m^2)$,  
is shown together in Fig. \ref{fig4}. 
In the classical spin flopping $m$ jumps from 0 to the line and 
the magnetization process is equivalent to the Ising model for $D<-1$.
In a way 
the transition at $H_{c3}$ is similar to the spin flopping 
because the transverse spin correlation has 
an exponential decay for LL2, while power-law one for LL1, 
which implies that even infinitesimal interchain interaction 
would yield the canted N\'eel order in LL1. 
LL2, however, 
is essentially produced by quantum effect, with no
classical correspondence. 

The quasi-1D antiferromagnet 
CsNiCl$_3$ can be a good candidate to test the transition at 
$H_{c3}$ because it is in the N\'eel phase in GS 
due to interchain interaction.\cite{csnicl} 
Thus 
it is interesting to consider the quasi-1D case. 
The interchain interaction enhances the most dominant spin 
correlation and leads to the long-range order. 
Thus it is expected that LL1 is changed into the canted N\'eel 
order, while LL2 into the incommensurate SDW with $k=(1-m)\pi$ 
along $H$. 
In addition the transition at $H_{c3}$ could become 
first-order one with a jump in the magnetization curve. 

\section{SUMMARY}

The finite cluster calculation and some size scaling analyses 
suggested that 
a field induced phase transition between two different Luttinger liquids
occurs at $H_{c3}$, between $H_{c1}$ and $H_{c2}$, in the 1D $S=1$ 
antiferromagnet with easy-axis anisotropy for 
$-1.49<D<-0.35$. 
In addition a re-entrant transition can appear 
for $-1.49<$D$<-1.32$. 
The transition is reminiscent to the spin flopping, 
but it 
gives rise to no
significant anomalous behavior in the magnetization curve. 

\section*{ACKNOWLEDGMENT}
We would like to thank Prof. M. Takahashi for  
fruitful discussions. 
We also thank 
the Supercomputer Center, Institute for
Solid State Physics, University of Tokyo for the facilities
and the use of the Fujitsu VPP500.

\end{document}